\documentstyle[preprint,aps,prd]{revtex}

\begin{document}
\draft

\title{ON THE STRUCTURE OF THE SCALAR MESONS \\
       $\bbox{f_0(975)}$ AND $\bbox{a_0(980)}$}
\author{G. Janssen \cite{Georg}}
\address{Department of Physics,
State University of New York at Stony Brook, \\
         Stony Brook, NY 11794, USA}
\author{B. C. Pearce}
\address{Department of Physics and Mathematical Physics, University of
         Adelaide,\\ Adelaide, S.A. 5005, Australia}

\author{K. Holinde and  J. Speth}
\address{Institut f\"{u}r Kernphysik, Forschungszentrum
         J\"{u}lich GmbH, \\ D-52425 J\"{u}lich, Germany}

\maketitle
\begin{abstract}
We investigate the structure of the scalar mesons $f_0(975)$ and
$a_0(980)$ within realistic meson-exchange models of the $\pi\pi$ and
$\pi\eta$ interactions. Starting from a modified version of the
J\"ulich model for $\pi\pi$ scattering we perform an analysis of the
pole structure of the resulting scattering amplitude and find, in
contrast to existing models, a somewhat large mass for the $f_0(975)$
($m_{f_0}=1015$ MeV, $\Gamma_{f_0}=30$ MeV).  It is shown that our
model provides a description of $J/\psi\rightarrow\phi\pi\pi/\phi KK$
data comparable in quality with those of alternative models.
Furthermore, the formalism developed for the $\pi\pi$ system is
consistently extended to the $\pi\eta$ interaction leading to a
description of the $a_0(980)$ as a dynamically generated threshold
effect (which is therefore neither a conventional $q\overline{q}$
state nor a $K\overline{K}$ bound state).  Exploring the corresponding
pole position the $a_0(980)$ is found to be rather broad
($m_{a_0}=991$ MeV, $\Gamma_{a_0}=202$ MeV).  The experimentally
observed smaller width results from the influence of the nearby
$K\overline{K}$ threshold on this pole.

\end{abstract}

\narrowtext

\section{Introduction}

With increasing experimental information about the different members
of the meson spectrum it becomes more and more important to develop a
consistent understanding of the observed mesons from a theoretical
point of view. For the low lying pseudoscalar, vector and tensor
mesons this has been done quite successfully within the framework of
the simple quark model assuming the mesons to be quark-antiquark
($q\overline{q}$) states grouped together into nonets.  For the scalar
mesons, however, several questions still remain to be answered, most
of them being related to the nature of the experimentally observed
mesons $f_0(975)$ and $a_0(980)$.

In a long standing controversial discussion the $f_0(975)$ has been
described as, for example, a conventional $q\overline{q}$ meson
\cite{Morgan74}, a $K\overline{K}$ molecule \cite{Weinstein90} or a
multiquark state \cite{Jaffe77}.  In order to discriminate between
different models Morgan and Pennington \cite{Morgan93a,Morgan93b}
recently investigated new data on the decay
$J/\psi\rightarrow\phi\pi\pi/\phi KK$ \cite{Falvard88,Lockman89}
concluding that a conventional Breit-Wigner (i.e.\ $q\overline{q}$)
structure is most probable for the $f_0(975)$. They claim that the
$\pi\pi/K\overline{K}$ amplitude is characterized by two poles near
the $K\overline{K}$ threshold on the $[bt]$ (second) and $[bb]$
(third) sheet (see Ref.\ \cite{Pearce89} and Sec.\ II for sheet
structure and notation). In the J\"ulich model of the $\pi\pi$
interaction \cite{Lohse90} the $f_0(975)$ appears to be a pure
$K\overline{K}$ bound state generated dynamically by vector-meson
exchange. Such a state has only one nearby pole on the $[bt]$ sheet, a
structure seemingly disfavored by the results of Ref.\
\cite{Morgan93a}.  However, it has been pointed out by Zou and Bugg
\cite{Zou94} that a model with only one nearby $[bt]$ sheet pole
(similar to a $K\overline{K}$ bound state) but a very broad $[bb]$
sheet pole is also compatible with the $J/\psi$ data. We have
therefore fitted this data with the $\pi\pi/K\overline{K}$ amplitude
resulting from our meson-exchange model and obtained a description
comparable in quality with those of the alternative models discussed
above.  This seems to support very recent developments \cite{BuggPriv}
that $J/\psi$ decay may no longer be regarded as the crucial
discriminant for the structure of the $f_0(975)$.

What turns out to be quite different in our analysis is the rather
large mass (interpreted as the pole position) of the $f_0(975)$
($m_{f_0}=1.015$ GeV, see Sec.\ III) which is a consequence of its
$K\overline{K}$ bound state nature in combination with a rather strong
coupling between $\pi\pi$ and $K\overline{K}$ channels.  While there
has been considerable discussion in the literature concerning the
width of the $f_0(975)$\cite{Morgan93b,Zou94,Zou93}, there seems to be
a consensus that the mass lies between 0.97 and 1.0 GeV.
Interestingly, our value for the mass is somewhat larger than this due
to the strong influence of the $K\overline{K}$ threshold.

Closely related to the $f_0(975)$ is the $a_0(980)$ which is observed
in the $\pi\eta$ channel. Both mesons couple to the $K\overline{K}$
channel and their masses lie very close to the $K\overline{K}$
threshold.  As with the $f_0(975)$, the nature of the $a_0(980)$ is
still under discussion, especially since the available experimental
information about $\pi\eta$ scattering is rather poor compared to
$\pi\pi$ scattering \cite{PDBa0}.  In Ref.\ \cite{Weinstein90} the
$f_0(975)$ and $a_0(980)$ emerge from a non-relativistic quark model
calculation to be degenerate $K\overline{K}$ bound states. However,
while the J\"ulich model of the $\pi\pi$ interaction comes to the same
conclusion for the $f_0(975)$ the situation turns out to be rather
different for the $a_0(980)$. The $K\overline{K}$ interaction
generated by vector-meson exchange, which for isospin $I=0$ is strong
enough to generate a bound state, is much weaker for $I=1$ making a
degeneracy of $a_0(980)$ and $f_0(975)$ impossible.

It is therefore one aim of this paper to address the question of
whether the close relation between $a_0(980)$ and $f_0(975)$ expressed
by their similar properties (masses, proximity to $K\overline{K}$
threshold, etc.) can be understood in the framework of our
meson-exchange model.  By extending the $\pi\pi$ model consistently to
the $\pi\eta$ system we find that this is indeed possible.  We obtain
a model where both the $a_0(980)$ and $f_0(975)$ originate from the
coupling to the $K\overline{K}$ channel.  However, the underlying
structure turns out to be quite different.  In contrast to the
$f_0(975)$, the $a_0(980)$ would not appear to be a $K\overline{K}$
bound state but a dynamically generated threshold effect with a
relatively broad pole on the $[bt]$ sheet.  We find poles responsible
for the observed $a_0(980)$ and $f_0(975)$ at complex energies
$(\mbox{Re} E, \mbox{Im} E) = (991,\pm101)$~MeV and $(1015,\pm15)$~MeV
respectively.  Despite such different pole positions, the observed
position and width of the structure in the cross sections turn out to
be rather similar.  This is directly attributable to the proximity of
the $K\overline{K}$ threshold at 991~MeV.

Section II contains the basic ingredients of our model for $\pi\pi$
and $\pi\eta$ scattering as well as the underlying formalism relevant
for the investigation of the $f_0(975)$ and $a_0(980)$. Results are
presented and discussed in Sec.\ III.  The paper ends with concluding
remarks in Sec.\ IV.

\section{Model}

In order to understand the underlying structure of the scalar mesons
$f_0(975)$ and $a_0(980)$ (referred to as $f_0$ and $a_0$ in the
following) we have to develop realistic models for $\pi\pi$ and
$\pi\eta$ scattering respectively.  The basis for the present work is
the J\"ulich meson exchange model for the $\pi\pi$ interaction whose
evaluation has been discussed in detail in former work
\cite{Lohse90,Pearce92b}.  Compared to the original version some
modifications have been performed which will be discussed in the
following.

The potentials for $\pi\pi\rightarrow\pi\pi$, $\pi\pi\rightarrow
K\overline{K}$ and $K\overline{K} \rightarrow K \overline{K}$ are
generated from the diagrams shown in Fig.\ \ref{figIIone}.  The figure
shows only $s$- and $t$-channel diagrams; $u$-channel processes
corresponding to the included $t$-channel processes are also included
when they contribute.  The scalar-isoscalar particle denoted by
$\epsilon$ in Fig.\ \ref{figIIone} effectively includes the singlet
and the octet member of the scalar nonet.  We investigated the effects
of $t$-channel $f_2(1270)$ and $\epsilon$ exchange but found their
effects to be negligible.  Hence, the results presented here do not
include them.  For completeness we have also added the $s$-channel
pole diagrams $\pi\pi/K\overline{K}\rightarrow \epsilon,\rho,f_2
\rightarrow\pi\pi/K\overline{K}$, which enable a unified description
of all partial waves.  However, the $s$-channel $\rho$ and $f_2$ poles
do not contribute to the description of the $f_0(975)$ or $a_0(980)$.

The coupling constant $g_{\rho\pi\pi}$, required for $t$- and
$u$-channel exchange diagrams, is determined from the decay widths of
the $\rho$.  The $s$-channel $\epsilon$, $\rho$ and $f_2$ exchanges
are renormalized by solving Eq.\ (\ref{scaeq}) below, so the bare
couplings and bare masses are adjusted to reproduce experimental data
in the appropriate partial waves.  All remaining coupling constants
are determined from SU(3) symmetry relations, and standard assumptions
about the octet/singlet mixing angles, as demonstrated in Ref.\
\cite{Lohse90} (see Tables \ref{tab:tparams} and \ref{tab:sparams}).

The scattering amplitudes are obtained by iterating these potentials
by using a coupled channel scattering equation.  We follow the
approach of \cite{Pearce92b} and use the Blankenbecler-Sugar formalism
(BbS, \cite{Blankenbecler66}) rather than time-ordered perturbation
theory \cite{Lohse90}.  The partial wave decomposed, coupled channel
scattering amplitude $T_{ij}$ [$i,j=\pi\pi(\pi\eta),K\overline{K}$;
indices for total angular momentum and isospin are suppressed] is
therefore given by
\begin{eqnarray}
T_{ij}(k', k; E) =&& V_{ij}(k',  k; E)
+ \sum_l \int dk'' {k''}^2 V_{il}(k', k''; E)
\nonumber \\
&&\times G_l(k'';E)T_{lj}(k'', k; E)  \nonumber \\
\label{scaeq}
\end{eqnarray}
where $E=\sqrt s$ denotes the total center of mass energy.

The BbS two-meson propagator for channel $l$ is denoted $G_l(k'';E)$
and can be written as
\begin{equation}
G_l(k'';E) = \frac{\omega_1+\omega_2}{(2\pi)^3  2 \omega_1\omega_2}
\frac{1}{(E^2-(\omega_1+\omega_2)^2)}
\end{equation}
where $\omega_1$ and $\omega_2$ are the on-mass shell energies for
particle $1$ and $2$ respectively,
$\omega_{\alpha}=\sqrt{k''^2+m_{\alpha}}$.  The on-shell momentum
$k_0$ for channel $l$ is defined by the singularity of $G_l$ to be

\begin{equation}
k_0 = \frac{\sqrt{(E^2 - (m_1 + m_2)^2) (E^2 - (m_1 - m_2)^2)}}{2E}
\label{eq:onshell}
\end{equation}

In order to determine the correct values for the mass and width of the
$f_0$ as they are predicted by our model, it is necessary to explore
the position of the poles of the scattering amplitude $T$. This can
only be done by taking into account the sheet structure of the
scattering amplitude which is imposed by the existence of $\pi\pi$ and
$K\overline{K}$ thresholds. We will use the notation of
\cite{Pearce89} for referring to the various sheets using a two
character string composed of the letters $t$ and $b$ to signify on
which sheet of each threshold the energy lies. The conventional second
sheet, for example, is therefore denoted by $[bt]$ indicating that the
energy is on the bottom sheet of the $\pi\pi$ channel and the top
sheet of the $K\overline{K}$ channel.  The sheet definition is
visualized in Fig.\ \ref{figIItwo}.

To determine the poles of the scattering amplitude $T_{ij}$ generated
by Eq.\ (\ref{scaeq}) we solve the following eigenvalue equation
\cite{Pearce84}

\begin{equation}
\sum_l [\lambda(E)\delta_{il}-V_{il}(E) G_l(E)] \phi_l(E) \equiv
\sum_l I_l(E) = 0,
\label{poleeq}
\end{equation}
where $\phi_l(E)$ denotes the eigenfunction for channel $l$, and
search for complex energies $E_R$ that result in an eigenvalue
$\lambda(E_R)=1$.

When the integral appearing in the momentum representation of Eq.\
(\ref{poleeq}) is performed, care must be taken to avoid the poles of
$G_l$ and to take into account the multi-sheet structure of $T_{ij}$.
Following the approach of Ref.\ \cite{Pearce89} this can be done for
physical (i.e., on the top sheet) energies $E_t$ simply by rotating
the contour of integration into the complex plane. For energies $E_b$
on the bottom sheet of threshold $l$ the residue of the pole of
$G_l(k'';E)$ at $k''=-k_0$ must be included. In summary we obtain
\cite{Pearce89}
\begin{eqnarray}
I_l(E_t) =&& \int_C dk'' {k''}^2 \bigl[
\lambda(E_t)\delta(k-k'')\delta_{il} \nonumber \\ && -
V_{il}(k,k'';E_t) G_l(k'';E_t)\bigr] \phi_l(k'';E_t)
\label{eq:poledettop}
\end{eqnarray}
and
\begin{eqnarray}
I_l(E_b) =&& \int_C dk'' {k''}^2
\bigl[\lambda(E_b)\delta(k-k'')\delta_{il} \nonumber \\ && -
V_{il}(k,k'';E_b) G_l(k'';E_b)\bigr] \phi_l(k'';E_b) \nonumber \\ && -
2\pi i \frac{(-k_0)}{4(2\pi)^3 E_b} V_{il}(k,-k_0;E_b)
\phi_l(-k_0;E_b)
\label{poledet}
\end{eqnarray}
where the contour $C$ is as shown in Fig.\ \ref{figIIthree}(a) for
$\mbox{Im}(E)\ge0$ and as in Fig.\ \ref{figIIthree}(b) for
$\mbox{Im}(E)<0$.

In practice, we implement this by using a Gaussian quadrature rule to
reduce Eqs.\ (\ref{poleeq})--(\ref{poledet}) to a matrix eigenvalue
equation, and then set $\lambda=1$ and search for energies such that
\begin{equation}
  \det[1 - V(E)G(E)] = 0 .
\end{equation}

In solving Eq.\ (\ref{poleeq}), problems may arise from the $t$- and
$u$-channel contributions to the potential $V_{il}$.  Inserting the
off-shell prescription of the BbS formalism (see Appendix A) we obtain
for the $t$-channel diagrams (and analogously for the $u$-channel)
\begin{eqnarray}
V_{il}(k',k;E) &\propto& [t-M^2]^{-1}\nonumber \\
&=&[-k^2-{k'}^2+2kk'\cos\theta
-M^2]^{-1}
\label{tchan}
\end{eqnarray}
where $M$ denotes the mass of the exchanged particle and $\theta$ is
the scattering angle.  For a fully off-shell potential this propagator
never becomes singular since $k$ and $k'$ vary along the same rotated
contour. This is not true, however, for the half off-shell potentials
appearing in Eq.\ (\ref{poledet}) which leads to a restriction on the
energy region that can be be searched using this method.  We have
calculated the locations of all these singularities in the complex
energy plane and found them to be far away from the energy region of
interest to this investigation.  This is also true for the $\pi\eta$
channel where the equivalent of Eq.\ (\ref{tchan}) looks more
complicated due to the unequal masses.

The formalism for calculating the $J/\psi\rightarrow\phi\pi\pi/\phi
KK$ mass spectra has been described in detail in Ref.\ \cite{Zou94}.
The input is the $\pi\pi/K\overline{K}$ amplitudes $T_{ij}$ which in
our model are generated by Eq.\ (\ref{scaeq}).  Our amplitudes are
normalized differently to those of Ref.\ \cite{Zou94}, with the
relation between the two being given by
\begin{equation}
T_{ij}^{\protect\cite{Zou94}} = [64\pi^2]^{-1} T_{ij} .
\end{equation}
Since the difference is just a constant factor it is absorbed in the
overall normalization of the $J/\psi$ mass spectra.  The amplitudes
for $J/\psi$ decay are given by

\begin{eqnarray}
F(\psi\rightarrow\phi\pi^+\pi^-) &=& \left [{\textstyle {2\over 3}}
\right]^{1/2} [\alpha_1(s)T_{11} + \alpha_2(s)T_{21}] , \nonumber \\
F(\psi\rightarrow\phi K^+K^-) &=& \left [{\textstyle{1\over 2}} \right
]^{1/2} [\alpha_1(s)T_{12} + \alpha_2(s)T_{22}] ,
\end{eqnarray}
with the real coupling functions $\alpha_i$ parameterized by
\begin{equation}
\alpha_i(s) = \gamma_{i0} + \gamma_{i1} s
\label{eq:params}
\end{equation}
where $\gamma_{i0}$ and $\gamma_{i1}$ are free parameters.  The
experimentally observed mass spectra are then formed from the modulus
squared of these amplitudes multiplied by phase space and an overall
normalization factor.

Finally, we develop our formalism for the $\pi\eta$ interaction in
order to obtain an understanding of the $a_0$ meson.  As for the
$f_0$, we include the coupling to the $K\overline{K}$ channel, where
the corresponding direct $K\overline{K}$ interaction is taken to be
exactly the same as for the $\pi\pi$ case, projected now onto isospin
$I=1$. In addition, our model includes the $t$-channel $K^*$ exchange
diagram shown in Fig.\ \ref{figIIfour} (and corresponding $u$-channel
diagram) which couples the $\pi\eta$ channel to $K\overline{K}$. The
spin-momentum structure of this potential is given by the same general
expression for the interaction between two pseudoscalar particles
which has been derived for the $\pi\pi$ interaction (see Ref.\
\cite{Lohse90} and Appendix A).  The potential corresponding to Fig.\
\ref{figIIfour} is obtained by using the appropriate coupling
constants and isospin factors (Tables \ref{tab:tparams} and
\ref{tab:isofactors}) and the BbS off-shell prescription for
different-mass particles (Appendix A).

Since the coupling constant $g_{\eta K K^*}$ is related by SU(3)
symmetry to $g_{\rho\pi\pi}$, the entire $\pi\eta$ $T$-matrix is
obtained with the addition of only one new parameter, the cutoff mass
$\Lambda_{\eta KK^*}$.

The $\pi\eta$ scattering amplitude $T$ and the location of its poles
follow from exactly the same formalism as was developed for the
$\pi\pi$ interaction.  Since the $\epsilon$ and $f_2$ $t$-channel
exchanges were found to be negligible, this model does not include any
direct $\pi\eta\rightarrow\pi\eta$ potential.

\section{Results and discussion}

Having specified all of the relevant formalism, we first look at
elastic $\pi\pi$ scattering where we essentially reproduce the results
of Ref.\ \cite{Lohse90} within our modified model.  Figure
\ref{figIIIone} shows the result of our fit to experimental data for
the $JI=00$ ($f_0$) partial wave as well as for $JI=02$, $11$, $20$
and $22$ with the complete set of parameters given in Tables
\ref{tab:tparams}--\ref{tab:barem}.  As could be expected from Refs.\
\cite{Lohse90,Pearce92b} we obtain quite good overall agreement with
the experimental situation.  The quality of the total fit obtained
with a relatively small number of parameters demonstrates the validity
of the model.  In particular we note that $t$-channel $\rho$ exchange
is the sole contributor to $JI=02$ and $22$, and, as can be seen in
the dotted curve in Fig.\ \ref{figIIIone}, provides a substantial part
of the low energy $JI=00$ interaction.  This suggests that the
spin-isospin structure provided by the $t$-channel meson exchanges
arising from a Lagrangian with effective meson degrees of freedom is
substantively correct.

In particular we are able to describe the structure appearing around
1.0 GeV in the isoscalar $\pi\pi$ S-wave which is assigned to the
$f_0$ meson. In our model this resonance like behavior is generated
dynamically by the strong attraction arising from $\rho$, $\omega$ and
$\phi$ exchange in the $K\overline{K}$ channel and we therefore do
{\it not} need a genuine scalar resonance with mass around 1.0 GeV.
This is demonstrated for $\delta_{00}$ by the dashed line in Fig.\
\ref{figIIIone} where we have excluded the $s$-channel $\epsilon$ pole
diagram. On the other hand, it is definitely necessary to include a
heavy scalar particle, namely the $\epsilon$ with mass around 1.4 GeV,
to describe the experimental data beyond 1.0 GeV (see the solid line
in Fig.\ \ref{figIIIone}).  This particle may be interpreted as the
member of the scalar nonet; i.e., it effectively parameterizes the
effect of both the isoscalar singlet and octet contributions and any
other higher mass states present in the scalar-isoscalar spectrum.
For smaller energies ($E\simeq 0.5$ GeV) the $\pi\pi$ S-wave is
characterized by a sizable phase shift dominantly generated by
$t$-channel $\rho$ exchange. The corresponding attraction between the
two pions forms a broad background to the $f_0$ (dotted curve in Fig.\
\ref{figIIIone}).

Having obtained a model which is able to describe experimental data on
elastic $\pi\pi$ scattering, we are now in a position to discuss the
pole positions of the scattering amplitude in the isoscalar $\pi\pi$
S-wave.

Looking at smaller energies first, we find a very broad pole on the
$[bt]$ sheet at complex energy $(\mbox{Re} E, \mbox{Im}
E)=(387,\pm305)$~MeV and its shadow-pole counterpart\cite{Pearce89} on
the $[bb]$ sheet at $(314,\pm428)$~MeV (see Table~\ref{tab:poles}).
At vanishing $\pi\pi/K\overline{K}$ coupling this pole is found on
both $K\overline{K}$ sheets at the same position but when the coupling
is increased they move apart.  This is demonstrated in Fig.\
\ref{figIIItwo} where we turned off the $\pi\pi/K\overline{K}$
coupling gradually, thus proving that we have really found a pole and
its corresponding shadow pole.  This shows that these poles are
generated by the strong $t$-channel $\rho$ exchange in the
$\pi\pi\rightarrow\pi\pi$ potential.  The pole closest to the physical
region, namely the one on $[bt]$, is the origin of the large $\pi\pi$
S-wave phase shifts below 1.0 GeV (see $\delta_{00}$ in Fig.\
\ref{figIIIone}).  Following Ref.\ \cite{Zou94} we denote it
$\sigma(400)$.

We find an additional pair of connected poles at lower energies
(Re($E$)$\simeq$500 MeV) on sheets $[tb]$ and $[bb]$ indicating that
this time the origin is the direct $K\overline{K}$
interaction. However, since both $[tb]$ and $[bb]$ poles are far away
from the physical region they do not have any effect on physical
observables and are therefore quite unimportant for the present
investigation.

Looking at higher energies, we find the poles generated by the
$\epsilon$ $s$-channel diagram necessary to describe data above 1.0
GeV. Including exclusively this diagram and turning off the
$\pi\pi/K\overline{K}$ coupling a pole is found at $(1354,\pm167)$~MeV
on $[bt]$ and $[bb]$ sheets (Fig.\ \ref{figIIIthree}, point A).  The
$t$-channel contributions to the $\pi\pi$ potential shift both poles
to point B of Fig.\ \ref{figIIIthree} before they move apart when the
$\pi\pi/K\overline{K}$ coupling is increased. Whereas the physically
unimportant $[bt]$ pole moves to very high energies the $[bb]$ pole is
finally found at $(1346,\pm249)$~MeV (see Table~\ref{tab:poles}). This
pole, lying closest to the physical region, defines the parameters of
the genuine scalar particle which effectively includes the singlet and
the octet member of the scalar nonet. We denote it by $f_0(1400)$
although it is more likely an effective parameterization of two scalar
resonances, such as $f_0(1400)$ and $f_0(1590)$. Both poles,
$\sigma(400)$ and $f_0(1400)$ form a background to the $f_0(975)$.  A
similar structure has been observed in Ref.\ \cite{Zou93}.

Next we investigate the energy region around the $K\overline{K}$
threshold and find a single pole on the $[bt]$ sheet at
$(1015,\pm15)$~MeV which clearly has to be assigned to the $f_0$ meson
and to the corresponding structure in $\pi\pi$ phase shifts and
inelasticities.  We therefore obtain
\begin{equation}
m_{f_0}=1015~{\rm MeV}; \hspace{2em} \Gamma_{f_0}=30~{\rm MeV} .
\end{equation}

In the zero $\pi\pi/K\overline{K}$ coupling limit the pole moves back
to $(985,0)$~MeV (i.e., on the real axis and below the $K\overline{K}$
threshold) and appears on sheets $[tt]$ and $[bt]$ (see Fig.\
\ref{figIIIfour}).  This demonstrates the bound state nature of the
$f_0$ within our model.

Compared to other models \cite{Morgan93a,Zou93} we find a rather high
mass (when interpreted as the pole position) for the $f_0$. This is a
consequence of the relatively strong coupling between the $\pi\pi$ and
$K\overline{K}$ channels which moves the pole from the real axis to
its final position. A similar movement of a bound state pole to
energies above the corresponding threshold was already found in Ref.\
\cite{Pearce89} and as the strength of the $\pi\pi/K\overline{K}$
coupling is constrained by the fit to experimental data, the
relatively high mass seems to be a more general feature of our bound
state model for the $f_0$.

The fact that the pole is on the $[bt]$ sheet and above the
$K\overline{K}$ threshold means that its effect will be seen most
strongly at the threshold.  This can be seen in, for example, the
partial wave cross section which is shown in Fig.\
\ref{fig:f0crosssection}.  The most evident feature is the strong dip
in the cross section which occurs precisely at the $K\overline{K}$
threshold.  There is no structure evident near the pole at
$(1015,\pm15)$~MeV.  Clearly it is insufficient to simply quote the
mass parameters of the $f_0(975)$ without describing in detail how the
threshold is incorporated since the two are inextricably linked.

For completeness, we finally look at the $\rho$ ($JI=11$) channel. As
could be expected we observe a pole on the $[bt]$ sheet and an
unimportant shadow pole on the $[bb]$ sheet with the position of the
former found at $(775,\pm82)$~MeV. The deviation from the usually
given Breit-Wigner parameters ($m_\rho=768.1$ MeV; $\Gamma_\rho/2=$76
MeV) is due to the existence of a non-pole background generated by the
$t$-channel diagrams.

So far we have discussed a model for $\pi\pi$ scattering which is able
to produce quite satisfactory agreement with experimental data and
leads to a $K\overline{K}$ bound state structure for the $f_0$ meson.
However, as was discussed in the introduction, it has been pointed out
by Morgan and Pennington \cite{Morgan93a,Morgan93b} that information
on elastic $\pi\pi$ scattering is not sufficient to discriminate
between alternative models for the structure of the $f_0$.  Whereas
they claim that the decay $J/\psi\rightarrow \phi\pi\pi/\phi
K\overline{K}$ demands two poles nearby the $K\overline{K}$ threshold
on sheets $[bt]$ and $[bb]$, Zou and Bugg \cite{Zou94} recently found
that a solution with only one nearby $[bt]$ pole is also compatible
with the experimental situation.  The latter model contains an
additional $[bb]$ sheet pole but since its position is far away from
$K\overline{K}$ threshold the pole structure is similar to those of
our $K\overline{K}$ bound state model.

Figure \ref{figIIIfive} shows the result of our fit to the $J/\psi$
decay data from DM2 \cite{Falvard88} and MK3 \cite{Lockman89}.  The
fit was obtained using
$\{\gamma_{10},\gamma_{11},\gamma_{20},\gamma_{21}\} =
\{7.716,-16.399,24.670,-9.418\}$ (see Eq.\ (\ref{eq:params}))
($\gamma_{10}$ and $\gamma_{20}$ are dimensionless while $\gamma_{11}$
and $\gamma_{21}$ are in GeV$^{-2}$).  The fit demonstrates that our
model is able to reproduce this data quite well.  Moreover, the
quality of the fit is strictly comparable in quality with those of
alternative models \cite{Morgan93b,Zou94} leading to the conclusion
that the bound state structure of the $f_0$ is not disfavored by the
$J/\psi$ criterion.  This finding is in agreement with a very recent
development calling into question the role of the $J/\psi$ decay as a
crucial discriminant between different models for the $f_0$
\cite{BuggPriv}.

Next we turn our attention to the $\pi\eta$ channel and the structure
of the $a_0$. It was already pointed out that we want to extend the
concepts applied to the $\pi\pi$ interaction {\it consistently} to the
$\pi\eta$ system; therefore, the $K\overline{K}$ interaction required
for a $\pi\eta/K\overline{K}$ coupled-channel approach is taken to be
exactly the same as for the $\pi\pi$ case (projected now onto isospin
$I=1$). However, as can be seen from Tables~\ref{tab:tparams} and
\ref{tab:isofactors}, the important $\rho$ exchange between two kaons
becomes repulsive for isospin $I$=1 destroying the $K\overline{K}$
bound state we found for $I$=0.  To illustrate this we considered only
the direct $K\overline{K}$ interaction and gradually changed the
isospin factor for $\rho$ exchange from $-3$ to 1.  Figure
\ref{figIIIsix} shows the motion of the bound state pole that results.
As the $K\overline{K}$ interaction decreases the bound-state pole on
the $[bt]$ (=$[tt]$) sheet crosses the $K\overline{K}$ threshold to
the $[tb]$ (=$[bb]$) sheet and moves down along the real axis to
energies far away from $K\overline{K}$ threshold.

Although the preceding remarks rule out a $K\overline{K}$ bound-state
(or anti-bound state) structure for the $a_0$ our consistent meson
exchange framework allows us to proceeded anyway to construct a model
for $\pi\eta$ scattering as described in Section II.  As already
noted, only one additional parameter is needed, namely the cutoff mass
$\Lambda_{\eta KK^*}$.  Before we discuss our results some remarks
should be made concerning the available experimental information
which, in contrast to the $\pi\pi$ interaction, is rather poor for the
$\pi\eta$ case. The parameters of the $a_0$ have been derived
indirectly from the mass spectra of relatively complicated processes
like $pp\rightarrow pp\eta\pi^+\pi^-$\cite{PDBa0} and recent values
obtained from conventional Breit-Wigner fits are
\begin{eqnarray}
m_{a_0} &=& (983\pm 2)~{\rm MeV}; ~ \Gamma_{a_0} = (57\pm 11)~{\rm
MeV}~~\cite{PDBa0}, \nonumber \\ m_{a_0} &=& (984\pm 4)~{\rm MeV}; ~
\Gamma_{a_0} = (95\pm 14)~{\rm MeV}~~\cite{Armstrong91}.
\label{a0para}
\end{eqnarray}
We note that there would still appear to be some uncertainty
concerning the width of the $a_0$.  Interestingly, the Particle Data
Group\cite{PDBa0} choose to exclude the results of Ref.\
\cite{Armstrong91} from their averaged width.  This seems surprising
since (a) it was derived from a larger event sample than any of the
values they did include, (b) it is from a more recent publication than
any value included, the most recent value included being ten years
older, and (c) they {\em did} include the results of Ref.\
\cite{Armstrong91} in their mass average.  As we will see, our results
favor the larger width obtained by Ref.\ \cite{Armstrong91}.

Figure \ref{figIIIseven}(a) shows the sensitivity of our calculation
for the $\pi\eta$ cross section on the only remaining relevant
parameter not constrained by the $\pi\pi$ system, namely the cutoff
mass $\Lambda_{\eta KK^*}$.

Although the shape of the cross section depends on the parameter
choice we always observe a peak around the $K\overline{K}$
threshold. Looking at the pole structure of the corresponding
$\pi\eta$ $T$-matrix a single pole on the $[bt]$ sheet is found for
each value of $\Lambda_{\eta KK^*}$ [Fig.\ \ref{figIIIseven}(b)].  For
a pole position below $K\overline{K}$ threshold the corresponding
cross sections have a rounded peak form whereas for values above
(where a $[bb]$ pole at the same position would be closer to the
physical region) we observe a cusp structure.

In order to get a feeling for the origin of the observed $[bt]$ pole
we have excluded the direct $K\overline{K}$ interaction in Fig.\
\ref{figIIIeight}.  Although the pole is now always found above
$K\overline{K}$ threshold it is still present.  We therefore conclude
it must originate from the transition potential $\pi\eta \rightarrow
K\overline{K}$ and hence is a dynamically generated effect of the
opening of the $K\overline{K}$ channel.  However, the direct
$K\overline{K}$ interaction appears to be an important contribution in
order to obtain reasonable agreement with experimental information. It
is obvious that the cusp-like cross sections of Fig.\
\ref{figIIIeight}(a) do not yield $a_0$ parameters in agreement with
the Breit-Wigner fits summarized in Eq.\ (\ref{a0para}).

Compared to $K^*$ exchange in the $\pi\pi\rightarrow K\overline{K}$
transition potential, the corresponding diagram for
$\pi\eta\rightarrow K\overline{K}$ appears to be stronger. This can
already be seen by calculating the factor $C$ of Eq.\ (\ref{Cfac}) for
$K^*$ exchange yielding $C(\pi\eta\rightarrow
K\overline{K})$=$\sqrt{2} C (\pi\pi\rightarrow K\overline{K})$.  In
addition, the total $\pi\pi/K\overline{K}$ transition potential is
reduced by a partial cancellation between $K^*$ $t$-channel and
$\epsilon$ $s$-channel exchange.  Figure \ref{figIIInine} demonstrates
the crucial role of the $\pi\eta\rightarrow K\overline{K}$ coupling
strength for the $a_0$ structure.  When this coupling is decreased
only slightly the $[bt]$ pole moves to an energy region where it no
longer has any effect on physical observables.

Figure \ref{figIIIten} shows the cross section for our full model with
$\Lambda_{\eta KK^*}$ adjusted so that a Breit-Wigner fit to our
calculated cross section roughly agrees with that obtained by fits to
the available data\cite{PDBa0,Armstrong91}.  The mass and width of the
$a_0$ determined from this cross section are $m_{a_0}\simeq 985~{\rm
MeV}$ and $\Gamma_{a_0}\simeq 110~{\rm MeV}$, in reasonable agreement
with the experimental values of Ref.\ \cite{Armstrong91}.  It is not
possible in our model to obtain the smaller width reported by Ref.\
\cite{PDBa0}.  The `real' values for the $a_0$ parameters, however,
can only be obtained from the position of the pole and so we finally
find in our model
\begin{equation}
m_{a_0} = 991~{\rm MeV}; \hspace{2em}
\Gamma_{a_0} = 202~{\rm MeV}.
\end{equation}

The $a_0$ width determined from our pole analysis turns out to be much
larger than the standard values [Eq.\ (\ref{a0para})].  The observed
narrower peak in the cross section is a consequence of the closeness
to the $K\overline{K}$ threshold.  This is in agreement with other
investigations \cite{Flatte76,Achasov80} which also find that the
$a_0$ is not really narrow.

\section{Conclusions}

In summary, we have investigated the structure of the scalar mesons
$f_0(975)$ and $a_0(980)$ in the framework of a meson exchange model
for $\pi\pi$ and $\pi\eta$ scattering. The latter has been obtained by
a consistent extension of the J\"ulich model for the $\pi\pi$
interaction \cite{Lohse90}.

Our solution for the $\pi\pi$ scattering amplitude is shown to be
compatible with the available experimental data sets. In particular,
we are able to describe data on $J/\psi\rightarrow \phi\pi\pi/\phi
K\overline{K}$ decay which has been considered to be an important
testing criterion for models of the $f_0(975)$.  By exploring the pole
positions of the $\pi\pi$ $T$-matrix (see Table~\ref{tab:poles}) we
found the $f_0(975)$ to be a narrow $K\overline{K}$ bound state of
relatively large mass ($m_{f_0}$= 1015 MeV).  We also found a very
broad $\sigma(400)$ pole on the $[bt]$ sheet (cf. Ref.\ \cite{Zou94})
which generates the large $\pi\pi$ S-wave phase shifts below 1.0 GeV.
For high energies, the pole of a heavy scalar particle with
$q\overline{q}$ structure is observed on the $[bb]$ sheet which is
interpreted as a mixture of two scalar mesons, such as $f_0(1400)$ and
$f_0(1590)$.

A consistent extension of our model to the $\pi\eta$ channel generates
a pole on the $[bt]$ sheet relatively close to $K\overline{K}$
threshold.  Looking at the corresponding $\pi\eta$ cross section we
find a peak structure in reasonable agreement with the experimental
information available for the parameters of the $a_0(980)$. However,
the mass and width derived from the pole position show the $a_0(980)$
to be rather broad.  We discussed the role of the direct
$K\overline{K}$ interaction for our results and found it to be not
essential for the generation of a pole though it improves the
agreement with experimental data. The origin of the $[bt]$ pole
assigned to the $a_0(980)$ is therefore the transition potential
$\pi\eta\rightarrow K\overline{K}$.

In conclusion, both scalar mesons, $f_0(975)$ and $a_0(980)$, result
from the coupling to the $K\overline{K}$ channel which explains in a
natural way their similar properties. The underlying structure,
however, is quite different.  Whereas the $f_0(975)$ appears to be a
$K\overline{K}$ bound state the $a_0(980)$ is found to be a
dynamically generated threshold effect.  To finally decide about the
parameters of these mesons additional experimental information on
$K\overline{K}$ production and the $\pi\eta$ system is required. A
corresponding experiment is, for example, proposed for the J\"ulich
proton synchrotron COSY\cite{COSY}.

\acknowledgments

This work was supported in part by Deutscher Akademischer
Austauschdienst (zweites Hochschulsonderprogramm) and the Australian
Research Council.  G. J. thanks G. E. Brown for the hospitality he
enjoyed in Stony Brook and for many useful discussions.

\appendix
\section{Potential Expressions}

In this appendix we give explicit expressions for the potential
describing the interaction between two pseudoscalar mesons of mass
$m_p$. Some of them have already been derived in \cite{Lohse90}.  We
write $V_{s/t}\equiv V_{s/t}(k_3,k_4;k_1,k_2;E)$ where $k_1-k_4$ are
the four-momenta of the external mesons and the subscript denotes an
$s$- or $t$-channel exchange.  We have
%{\renewcommand{\theenumi}{\alph{enumi}}
\begin{description}
\item[\rm Scalar-meson exchange:]\mbox{}
\begin{eqnarray}
{\cal L}_{pps} &=& \frac{g_{pps}}{m_p} \partial^\mu\phi_p
  \partial_\mu\phi_p \phi_s \nonumber \\ V_t &=& \frac{4C}{m_p^2}
  \frac{k_{1\mu}k_3^\mu k_{2\nu}k_4^\nu}{t-M^2} \nonumber \\ V_s &=&
  \frac{4C}{m_p^2} \frac{k_{1\mu}k_2^\mu k_{3\nu}k_4^\nu}{s-M_0^2}
\label{eq:scalar}
\end{eqnarray}

\item[\rm Vector-meson exchange:]\mbox{}
\begin{eqnarray}
{\cal L}_{ppv} &=& g_{ppv} \phi_p \partial_\mu\phi_p \phi^\mu_v
\nonumber \\ V_t &=& C \sum_{\lambda} \frac{(k_1+k_3)^\mu(k_2+k_4)^\nu
\epsilon^*_\mu \epsilon_\nu}{t-M^2} \nonumber \\ V_s &=& C
\sum_{\lambda} \frac{(k_1-k_2)^\mu(k_3-k_4)^\nu \epsilon^*_\mu
\epsilon_\nu}{s-M_0^2}
\label{eq:vector}
\end{eqnarray}

\item[\rm Tensor-meson exchange:]\mbox{}
\begin{eqnarray}
{\cal L}_{ppt} &=& \frac{g_{ppt}}{m_p} \partial^\mu\phi_p \partial^\nu
  \phi_p \phi_{t\mu\nu} \nonumber \\ V_t &=& \frac{C}{m_p^2}
  \sum_\lambda (k_{1\mu}k_{3\nu} + k_{3\mu}k_{1\nu})
  (k_{2\sigma}k_{4\tau} + k_{4\sigma}k_{2\tau}) \nonumber \\ &&\times
  \frac{\epsilon^{*\mu\nu} \epsilon^{\sigma\tau}}{t-M^2} \nonumber \\
  V_s &=& \frac{C}{m_p^2} \sum_\lambda (k_{1\mu}k_{2\nu} +
  k_{2\mu}k_{1\nu}) (k_{3\sigma}k_{4\tau} + k_{4\sigma}k_{3\tau})
  \nonumber \\ &&\times \frac{\epsilon^{*\mu\nu}
  \epsilon^{\sigma\tau}}{s-M_0^2}
\label{eq:tensor}
\end{eqnarray}
\end{description}
Here, $\epsilon^{\mu}$ denotes the polarization vector of an exchanged
vector meson, which depends on the exchanged momentum and helicity
$\lambda$, while $\epsilon^{\mu\nu}$ is the corresponding object for
exchanged tensor mesons.  The mass of the exchanged particle is
denoted by $M$.  For $s$-channel processes the multiple scattering
series renormalizes the exchanged meson so a bare mass $M_0$ is used,
which in each case is adjusted to fit an observed resonance.

The $u$-channel contributions can be obtained by exchanging $k_3$ and
$k_4$ and replacing $t$ by $u$.

The overall factor $C$ is given by
\begin{equation}
C = g^2 f n F^2.
\label{Cfac}
\end{equation}
Here, $F$ denotes a standard form factor of dipole type
\begin{eqnarray}
\mbox{$t$-, $u$-channel:} \hspace{2em} && F=\left(
        \frac{2\Lambda^2-M^2}{2\Lambda^2-({\bf k'}\mp{\bf k})^2}
        \right)^2 \nonumber \\ \mbox{$s$-channel:} \hspace{2em} &&
        F=\left( \frac{2\Lambda^2-k_0^2(M)}{2\Lambda^2-|{\bf k}|^2}
        \right)^2
\end{eqnarray}
where $\bf k$ and $\bf k'$ denote the initial and final c.m.\ momenta
and $k_0(E)$ is the on-shell momentum defined in Eq.\
(\ref{eq:onshell}).  [For $s$-channel, the $F^2$ appearing in Eq.\
(\ref{Cfac}) is really $F({\bf k})F({\bf k'})$.]  The corresponding
cutoff masses $\Lambda$, together with coupling constants $g^2/4\pi$,
isospin factors $f$ and bare masses $M_0$, are given in
Tables~\ref{tab:tparams}--\ref{tab:isofactors}.  The isospin factors
$f$ are derived from the standard isospin structure of the
Lagrangians, which was suppressed in Eqs.\
(\ref{eq:scalar})--(\ref{eq:tensor}).  The factor $n$ takes into
account the normalization of states for identical particles
\cite{Lohse90}.  It is $1/2$ for $\pi\pi\rightarrow\pi\pi$, $1/\sqrt
2$ for $\pi\pi\rightarrow K\overline{K}$ and $1$ for
$K\overline{K}\rightarrow K\overline{K}$, $\pi\eta\rightarrow\pi\eta$
and $\pi\eta\rightarrow K\overline{K}$.

To evaluate the potentials explicitly the BbS off-shell prescription
has to be applied for the different momenta;

\begin{equation}
k_1=\left( \frac{\sqrt s}{2} + c, {\bf k} \right); \hspace{2em}
k_2=\left( \frac{\sqrt s}{2} - c,-{\bf k} \right),
\end{equation}
and
\begin{equation}
k_3=\left( \frac{\sqrt s}{2} + c', {\bf k'} \right); \hspace{2em}
k_4=\left( \frac{\sqrt s}{2} - c',-{\bf k'} \right),
\end{equation}
where
\begin{equation}
c,c' = \left \{ \begin{array} {cl} {m_\pi^2-m_\eta^2\over 2\sqrt s_{
                    }} & \mbox{for $\pi\eta$} \\ 0 & \mbox{for
                    $\pi\pi$, $K\overline{K}$} .  \end{array} \right.
\end{equation}

Finally, a standard partial wave expansion is performed to obtain
potentials suitable for use in Eq.\ (\ref{scaeq}).

%\bibliography{References}
%\bibliographystyle{prsty}

\begin{figure}
\caption{Diagrams included in the potentials
$\pi\pi\rightarrow\pi\pi$, $\pi\pi\rightarrow K\overline{K}$ and
$K\overline{K}\rightarrow K\overline{K}$.}
\label{figIIone}
\end{figure}

\begin{figure}
\caption{Sheets of the energy plane and their labeling.  In
 parentheses we have given the conventional notation.  The double
 lines symbolize $\pi\pi$ and $K\overline{K}$ unitarity cuts.}
\label{figIItwo}
\end{figure}

\begin{figure}
\caption{Rotated integration contours used for evaluating the
integrals of Eqs.\ (\protect\ref{eq:poledettop}) and
(\protect\ref{poledet}).}
\label{figIIthree}
\end{figure}

\begin{figure}
\caption{Diagram included in the potential $\pi\eta\rightarrow
K\overline{K}$.}
\label{figIIfour}
\end{figure}

\begin{figure}
\caption{Results of the modified J\"ulich $\pi\pi$ interaction model.
Data are from Refs.\ \protect\cite{Frogatt77,Martin76,Ochs73}. The
solid line shows the result of our full model. For the dashed line in
$\delta_{00}$ we have excluded the $\epsilon$ $s$-channel pole
diagram. The dotted line contains only $t$- and $u$-channel $\rho$
exchange.}
\label{figIIIone}
\end{figure}

\begin{figure}
\caption{The positions of the $\sigma(400)$ pole on $[bt]$ and its
shadow pole on $[bb]$ for our full model (squares). For decreasing
$\pi\pi/K\overline{K}$ coupling they move along the curves indicated
and the dot gives their position in the zero coupling limit.}
\label{figIIItwo}
\end{figure}

\begin{figure}
\caption{The $f_0(1400)$ pole on $[bb]$ and its shadow pole on $[bt]$.
Notations are as in Fig.\ \protect\ref{figIIItwo}. For point A, only
the diagram $\pi\pi\rightarrow\epsilon\rightarrow\pi\pi$ is considered
while point B includes in addition the $t$-channel diagrams for
$\pi\pi\rightarrow\pi\pi$.}
\label{figIIIthree}
\end{figure}

\begin{figure}
\caption{The $f_0(975)$ pole on $[bt]$. Notations as in Fig.\
\protect\ref{figIIItwo}.}
\label{figIIIfour}
\end{figure}

\begin{figure}
\caption{The calculated $JI=00$ $\pi\pi$ cross section.}
\label{fig:f0crosssection}
\end{figure}
\begin{figure}
\caption{Fit to $J/\psi\rightarrow \phi\pi\pi/\phi KK$ data.  The
upper panel shows the fit to data of Ref.\ \protect\cite{Lockman89}
and the lower panel to Ref.\ \protect\cite{Falvard88}.}
\label{figIIIfive}
\end{figure}

\begin{figure}
\caption{The movement of the $K\overline{K}$ bound-state pole as the
strength of $\rho$ exchange decreases (finally becoming repulsive).
Starting point is the attractive interaction corresponding to the
$\pi\pi$ system where $I=0$ (isospin factor $f=-3$) which gives rise
to the bound state (dot).}
\label{figIIIsix}
\end{figure}

\begin{figure}
\caption{The $\pi\eta$ cross section (a) and the corresponding pole
position (b) for different choices of the parameter $\Lambda_{\eta
KK^*}$.  A: 3.6 GeV; B: 3.1 GeV; C: 2.6 GeV.}
\label{figIIIseven}
\end{figure}

\begin{figure}
\caption{The same as Fig.\ \protect\ref{figIIIseven} but without any
direct $K\overline{K}$ interaction.  The values of $\Lambda_{\eta
KK^*}$ are A: 5.0 GeV; B: 4.0 GeV; C: 3.1 GeV.}
\label{figIIIeight}
\end{figure}

\begin{figure}
\caption{The movement of the $a_0(980)$ pole as the
$\pi\eta/K\overline{K}$ coupling strength is decreased. When the pole
crosses the real axis the reduction factor is about 0.8.}
\label{figIIInine}
\end{figure}

\begin{figure}
\caption{The $\pi\eta$ cross section for our full model.}
\label{figIIIten}
\end{figure}
\newpage

{ \renewcommand{\arraystretch}{1.4} \narrowtext
\begin{table}
\caption{Vertex parameters for $t$-channel exchanges.  Relations
between coupling constants are obtained using SU(3) and ideal mixing
between the octet and singlet.}
\label{tab:tparams}
\begin{tabular}{cdr}
Vertex & $g$ & \multicolumn{1}{c}{$\Lambda$ [MeV]} \\ \hline
$\pi\pi\rho$ & 6.04 & 1355 \\ $\pi KK^*$ & $g_{\pi KK^*} = g_{\pi
\overline{K}\overline{K^*}} = -\frac{1}{2}g_{\pi\pi\rho}$ & 1900 \\
$KK\rho$ & $g_{KK\rho} = g_{\overline{K}\overline{K}\rho} =
\frac{1}{2}g_{\pi\pi\rho}$ & 1850 \\ $KK\omega$ & $g_{KK\omega} =
-g_{\overline{K}\overline{K}\omega} = \frac{1}{2}g_{\pi\pi\rho}$ &
2800 \\ $KK\phi$ & $g_{KK\phi} = -g_{\overline{K}\overline{K}\phi} =
\frac{1}{\sqrt{2}}g_{\pi\pi\rho}$ & 2800 \\ $\eta KK^*$ & $g_{\eta
KK^*} = -g_{\eta \overline{K}\overline{K^*}} =
-\frac{\sqrt{3}}{2}g_{\pi\pi\rho}$ & 3290
\end{tabular}
\end{table}
}

{
\narrowtext
\begin{table}
\caption{Vertex parameters for $s$-channel exchanges.  The exchanged
meson is identified with a superscript $(0)$ since it is a bare meson.
The $\epsilon^{(0)}$ contributes to the description of the
$f_{0}(975)$.  None of the other parameters contribute to $f_{0}$ or
$a_{0}$.}
\label{tab:sparams}
\begin{tabular}{cdr}
Vertex & $g$ & \multicolumn{1}{c}{$\Lambda$ [MeV]} \\ \hline
$\pi\pi\epsilon^{(0)}$ & 0.286 & 925 \\ $KK\epsilon^{(0)}$ &
$-$0.286\tablenote{Since the singlet/octet mixing angle for the scalar
nonet is not known, $g_{KK\epsilon^{(0)}}$ is a free parameter.}  &
1200 \\ $\pi\pi\rho^{(0)}$ & 5.32 & 1647 \\ $KK\rho^{(0)}$ &
$\frac{1}{2} g_{\pi\pi\rho^{(0)}}$ & 830 \\ $\pi\pi f_2^{(0)}$ & 1.23
& 995 \\ $KK f_2^{(0)}$ & $\frac{2}{3} g_{\pi\pi f_2^{(0)}}$ & 937
\end{tabular}
\end{table}
}

{
\narrowtext
\begin{table}
\caption{Bare masses $M_0$ used in the $s$-channel exchanges, in MeV.}
\label{tab:barem}
\begin{tabular}{ccc}
$\epsilon^{(0)}$ & $\rho^{(0)}$ & $f_2^{(0)}$ \\
\hline
1520             & 1125         & 1660
\end{tabular}
\end{table}
}

{
\narrowtext
\begin{table}
\caption{Isospin factors for each meson exchange diagram used.}
\label{tab:isofactors}
\begin{tabular}{cccccc}
Potential & Meson & Type & \multicolumn{3}{c}{Factor $f$} \\ & & &
          $I=0$ & $I=1$ & $I=2$ \\ \hline $\pi\pi\rightarrow\pi\pi$ &
          $\rho$ & $t$-channel & $-$2 & $-$1 & 1 \\ & $\epsilon^{(0)}$
          & $s$-channel & 3 & 0 & 0 \\ & $\rho^{(0)}$ & $s$-channel &
          0 & 2 & 0 \\ & $f_2^{(0)}$ & $s$-channel & 3 & 0 & 0
          \\[\medskipamount] $\pi\pi\rightarrow K\overline{K}$ & $K^*$
          & $t$-channel & $\sqrt{6}$ & 2 & --- \\ & $\epsilon^{(0)}$ &
          $s$-channel & $-\sqrt{6}$ & 0 & --- \\ & $\rho^{(0)}$ &
          $s$-channel & 0 & 2 & --- \\ & $f_2^{(0)}$ & $s$-channel &
          $-\sqrt{6}$ & 0 & --- \\[\medskipamount]
          $K\overline{K}\rightarrow K\overline{K}$ & $\rho$ &
          $t$-channel & $-$3 & 1 & --- \\ & $\omega$ & $t$-channel & 1
          & 1 & --- \\ & $\phi$ & $t$-channel & 1 & 1 & --- \\ &
          $\epsilon^{(0)}$ & $s$-channel & 2 & 0 & --- \\ &
          $\rho^{(0)}$ & $s$-channel & 0 & 2 & --- \\ & $f_2^{(0)}$ &
          $s$-channel & 2 & 0 & --- \\[\medskipamount]
          $\pi\eta\rightarrow K\overline{K}$ & $K^*$ & $t$-channel &
          --- & $\sqrt{2}$ & ---
\end{tabular}
\end{table}
}

{
\narrowtext
\begin{table}
\caption{A summary of all poles found in the
$\pi\pi-K\overline{K}-\pi\eta$ system.}
\label{tab:poles}
\begin{tabular}{cclcl}
I & J & Sheet & Pole position & Comment \\ & & & [MeV] & \\ \hline 0 &
  0 & $[bt]$ (II) & $(387,\pm305)$ & $\sigma(400)$ \\ 0 & 0 & $[bb]$
  (III) & $(314,\pm428)$ & $\sigma(400)$ shadow pole \\ 0 & 0 & $[bt]$
  (II) & $(1015,\pm15)$ & $f_0(975)$ \\ 0 & 0 & $[bb]$ (III) &
  $(1346,\pm249)$ & effective $f_0(1400)-f_0(1590)$ \\ 1 & 1 & $[bt]$
  (II) & $(775,\pm82)$ & $\rho$ \\ 1 & 0 & $[bt]$ (II) &
  $(991,\pm101)$ & $a_0(980)$
\end{tabular}
\end{table}
}

\end{document}